\documentstyle[prl,multicol,aps,epsfig]{revtex}
\renewcommand{\widetext}
{\end{multicols}\global\columnwidth42.5pc}
\begin{document}
\newcommand{\be}{\begin{equation}}
\newcommand{\ee}{\end{equation}}
\newcommand{\bea}{\begin{eqnarray}}
\newcommand{\eea}{\end{eqnarray}}
\newcommand{\br}{{\bf r}}
\newcommand{\bk}{{\bf k}}
\newcommand{\bq}{{\bf q}}
\newcommand{\bn}{{\bf n}}
\newcommand{\bp}{{\bf p}}
\newcommand{\ve}{\varepsilon}

\draft
\title{Cyclotron resonance harmonics in the ac response of a 2D electron
gas with smooth disorder}
\author{I.A.~Dmitriev$^{1,*}$, A.D.~Mirlin$^{1,2,\dagger}$, and
D.G.~Polyakov$^{1,*}$ }
\address{$^1$Institut
f\"ur Nanotechnologie, Forschungszentrum Karlsruhe, 76021 Karlsruhe,
Germany}
\address{$^2$Institut f\"ur Theorie der Kondensierten Materie,
Universit\"at Karlsruhe, 76128 Karlsruhe, Germany}
\date{\today}
\maketitle
\begin{abstract}
The frequency-dependent conductivity $\sigma_{xx}(\omega)$ of 2D
electrons subjected to a transverse magnetic field and smooth disorder
is calculated. The interplay of Landau quantization and disorder
scattering gives rise to an oscillatory structure that survives in the
high-temperature limit. The relation to recent experiments on
photoconductivity by Zudov {\it et al.} and Mani {\it et al.} is discussed.
\end{abstract}
\begin{multicols}{2}
\narrowtext

The magnetotransport properties of a high-mobility 2D electron gas
(2DEG) in semiconductor heterostructures are of great importance from
the point of view of both fundamental physics and
applications. Important information about the dynamical and spectral
properties of the system is provided by its response to a microwave
field. Within the quasiclassical Boltzmann theory, the dissipative
{\it ac} conductivity
$\sigma_{xx}(\omega)=\sigma_+(\omega)+\sigma_-(\omega)$ of a
non-interacting 2DEG in a magnetic field $B$ is given by the Drude
formula (we neglect spin for simplicity),
\be
\label{drude}
\sigma^{(D)}_{\pm}(\omega) = {e^2\nu_0 v_F^2\tau_{\rm tr,0}\over 4
  [1+(\omega_c\pm\omega)^2\tau_{\rm tr,0}^2]}~, 
\ee 
where $v_F$ is the Fermi velocity, $\nu_0=m/2\pi$ (with $\hbar=1$) the
density of states (DOS), $\tau_{\rm tr,0}$ the transport relaxation
time at $B=0$, $\omega_c=eB/mc$ the cyclotron frequency, and $m$ is
the electron effective mass.  For a sufficiently clean sample,
$\omega\tau_{\rm tr,0}\gg 1$, Eq.~(\ref{drude}) predicts a sharp
cyclotron resonance (CR) peak at $\omega_c=\omega$.

It has been shown by Ando \cite{ando75,afs} that the Landau
quantization of the orbital electron motion leads, in combination with
disorder, to the emergence of harmonics of the CR at
$\omega=n\omega_c$, $n=2,3,\ldots$.  Indeed, such a structure was
experimentally observed \cite{abstreiter76}.  The analytical
calculations of Ref.~\cite{ando75} were performed, however, only for
fully separated Landau levels with point-like scatterers
\cite{footAndo}.

Very recently, great interest in the transport properties of a 2DEG in
a microwave field has been caused by experiments on photoconductivity
of exceptionally-high-mobility samples by Zudov {\it et al.}
\cite{zudov} and Mani {\it et al.} \cite{mani}, where oscillations
controlled by the ratio $\omega/\omega_c$ were observed.  Remarkably,
these oscillations persisted down to magnetic fields as low as $B\sim
10\: {\rm mT}$, an order of magnitude smaller than the field at which
the Shubnikov-de Haas oscillations were damped.  The experiments
\cite{zudov,mani} triggered an outbreak of theoretical activity. Durst
{\it et al.} \cite{durst} proposed (see also
Refs.~\cite{anderson,shi}) that the oscillations are governed by the
following mechanism: an electron is excited by absorbing a photon with
energy $\omega$ close to $n\omega_c$ and is scattered by disorder. In
view of the oscillatory structure of the DOS, this leads to an extra
contribution to the {\it dc} conductivity. In fact, a very similar
mechanism of oscillatory photoconductivity was proposed long ago
\cite{ryzhii} for the case of a strong {\it dc} electric field.

While the proposal of Ref.~\cite{durst} is very appealing,
calculations presented there involve a number of assumptions and
approximations, which complicates a comparison with experiment. First,
the consideration of Ref.~\cite{durst} is restricted to the case of
white-noise disorder with $\tau_{\rm tr,0}=\tau_{\rm s,0}$, where
$\tau_{s,0}$ is the single-particle relaxation time at $B=0$. On the
other hand, the experiments are performed on high-mobility samples
with smooth disorder, $\tau_{\rm tr,0}/\tau_{\rm s,0}\simeq 50$.
Second, Ref.~\cite{durst} neglects all vertex corrections and discards
inelastic processes. As we argue below, the inelastic relaxation rate
is of central importance for the magnitude of the photoconductivity.

The development of the full theory of the oscillatory
photoconductivity remains thus a challenging task, which we postpone
to a future work. In this paper we address the problem of the {\it ac}
response of a 2DEG with smooth disorder. On top of fundamental
theoretical importance and experimental relevance this problem
possesses on its own, it is closely related to the photoconductivity
mechanism considered above. Indeed, the key ingredient of this
mechanism---absorption of a photon---is governed by the dissipative
{\it ac} conductivity $\sigma_{xx}(\omega)$. We will return to this
relation in the end of the paper, where we discuss implications of our
results for the photoconductivity.

We consider a 2DEG subjected to magnetic field $B$ and a random potential
$U({\bf r})$ characterized by a correlation function $\langle
U(\br)U(\br')\rangle = W(|\br-\br'|)$ of a spatial range $d$. The total and
the transport relaxation rates induced by disorder at $B=0$ are given by
\bea\nonumber
\label{rates}
\left.
\begin{array}{l}
\tau_{\rm s,0}^{-1} \\ 
\tau_{\rm tr,0}^{-1}
\end{array}
\right\}
= 2\pi\nu_0\int{d\phi\over 2\pi} \,\tilde{W}(2k_F\sin{\phi\over 2})\times
\left\{
\begin{array}{l}
1\\ 
(1-\cos\phi)
\end{array}
\right. , \eea 
where $\tilde{W}(\bq)$ is the Fourier transform of $W(\br)$.  While we
are mainly interested in the experimentally relevant case of smooth
disorder, when impurities are separated from the 2DEG by a spacer of
width $d\gg k_F^{-1}$, with $\tau_{\rm tr,0}/\tau_{\rm s,0}\sim (k_F
d)^2\gg 1$, our results are valid for arbitrary $d$ (including
short-range disorder with $\tau_{\rm tr,0}/\tau_{\rm s,0}\sim 1$).

To calculate (the real part of) the conductivity, we use the Kubo formula,
\bea \nonumber \sigma_{xx}(\omega)&=&-\,\frac{e^2}{4\pi
  V}\int\limits^\infty_{-\infty}\!
{{d\ve}\over{\omega}}\,(f_\ve-f_{\ve+\omega})\\
&\times&{\rm Tr}\:\overline{ \hat{v}_x(G^A_{\ve+\omega}-G^R_{\ve+\omega})
  \hat{v}_x(G^A_{\ve}-G^R_{\ve})}~,
\label{sigma1}
\eea 
where $f_\ve$ is the Fermi distribution, $G^{R,A}$ are the retarded
and advanced Green functions, the bar denotes impurity averaging, and
$V$ is the system area. We will treat disorder within the
self-consistent Born approximation (SCBA) \cite{afs}, which is
justified provided the disorder correlation length satisfies $d\ll
l_B$ and $d\ll v_F\tau_{\rm s,0}$, where $l_B=(c/eB)^{1/2}$ is the
magnetic length \cite{raikh}. The Green function in the Landau level
(LL) representation, $G_n^R = (G_n^A)^*$, is given by the SCBA
equations \cite{afs,raikh},
\be
G^{R}_n(\ve)=(\ve-\ve_n-\Sigma_\ve)^{-1},\;\;
\Sigma_\ve=\frac{\omega_c}{2\pi\tau_{s,0}}\sum\limits_n G^{R}_n(\ve)~,
\label{born}
\ee
where $\ve_n=(n+{{1}\over{2}})\omega_c$ is the n-th LL energy (Fig.~1a).

\begin{figure}
\narrowtext
\centerline{ {\epsfxsize=8.5cm{\epsfbox{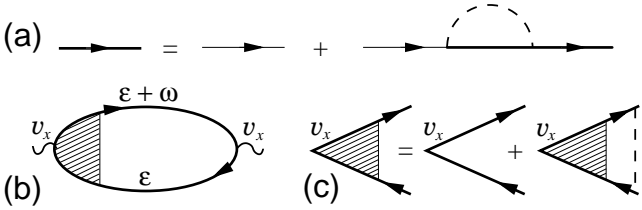}} }}
\vspace{3mm}
\caption{(a) SCBA equation for the Green function;\\
(b) the dynamical conductivity with vertex correction (c).}
\label{fig1}
\end{figure}

We will assume throughout the paper that $\omega,\omega_c \ll \ve_F$,
so that the relevant LL indices are large, $n\simeq \ve_F/\omega_c\gg
1$. Further, we will concentrate on the regime of strongly overlapping
LLs, $\omega_c\tau_{s,0}/\pi \ll 1$; the opposite case will be briefly
discussed in the end. To evaluate the self-energy in Eq.~(\ref{born}),
we use the Poisson   formula, $\sum_n F_n=\sum_k\int dx F(x)\exp (2\pi
i k x)$. The $k=0$ term yields then the $B=0$ result,
$\Sigma^{(0)}=i/2\tau_{s,0}$, while the $k=\pm 1$ contributions
provide the leading correction, 
\be
\Sigma^{(1)}_\ve=(i/2\tau_{s,0})\left[\,1-2\delta \exp(-2\pi
i\ve/\omega_c)\,\right]~, \label{self} 
\ee 
with $\delta=\exp(-\pi/\omega_c \tau_{s,0})$ serving as a small
parameter of the expansion. According to Eqs.~(\ref{born}),
(\ref{self}), the oscillatory correction to the DOS due to the LL
quantization reads 
\be
\Delta\nu(\ve)/\nu_0=-2\delta\cos(2\pi\ve/\omega_c)~.  \label{dens}
\ee

The  conductivity  (\ref{sigma1})  is  given  diagrammatically  by  an
electronic bubble  with a vertex correction,  i.e., by a  sum of
ladder diagrams,  Fig.~1b,c.  We evaluate  first  the  bare bubble
 $\sigma^b$ (which is  sufficient for the  case of white-noise
 disorder).  Making use of velocity  matrix elements in the LL
 representation, we get for $\ve_F\gg \omega,\omega_c$:
\bea &&\sigma_\pm^b(\omega)=\frac{e^2 v_F^2\nu_0}{4}\!\int\!
{{d\ve}\over{\omega}}\,(f_\ve-f_{\ve+\omega})\,{\rm
  Re}\,(\Pi_\pm^{RA}-\Pi_\pm^{RR})~,
\label{sigma2}\\
&&\Pi_\pm^{RR(RA)} = {\omega_c\over 2\pi}
\sum_n G^R_{n\pm 1}(\ve+\omega)G^{R(A)}_n(\ve)~.
\label{bubble}
\eea
Using again the Poisson formula, we find
\bea
&&\Pi_\pm^{RR}=\frac{\tau_{s,0}(\Sigma^{(1)}_\ve-\Sigma^{(1)}_{\ve+\omega})}
{\omega\pm\omega_c+\Sigma^{(1)}_\ve-\Sigma^{(1)}_{\ve+\omega}}~,
\nonumber \\
&&\Pi_\pm^{RA}=\frac{\tau_{s,0}(\Sigma^{(1)*}_\ve-\Sigma^{(1)}_{\ve+\omega})}
{\omega\pm\omega_c+\Sigma^{(1)*}_\ve-\Sigma^{(1)}_{\ve+\omega}}~.
\label{Pi}
\eea

For the case of smooth disorder we have to take into account the
vertex correction (Fig.~1c) while averaging in
Eq.~(\ref{sigma1}). This is a non-trivial task since the disorder
mixes strongly the LLs, thus seriously complicating a direct
calculation in the LL representation. We choose instead a different
way, which is suggested by the quasiclassical nature of the problem,
$\ve_F\tau_{s,0}\gg1$. It is instructive to recall first how the
vertex correction is calculated at $B=0$. The vertex function
$\Gamma_\pm^{RA}(\br_1,\br_2)$ [the average of $G^R v_\pm G^A$ with
$v_\pm=(v_x\pm iv_y)/2$] depends then on $\br_1-\br'_2$ only, yielding
$\Gamma_\pm^{RA}(\bp)$ in the Fourier space. In the quasiclassical
regime the momentum integrals are dominated by the vicinity of the
Fermi surface, reducing $\Gamma_\pm^{RA}(\bp)$ to
$\Gamma_\pm^{RA}(\phi)$, where $\phi$ is the polar angle of velocity
on the Fermi surface. The equation for $\Gamma_\pm^{RA}(\phi)$ is then
easily solved, yielding $\Gamma_\pm^{RA}(\phi)=(v_F/2)e^{\pm i\phi}
\tau_{\rm tr,0}/\tau_{\rm   s,0}$.

We are now going to generalize this quasiclassical calculation onto
the case of a finite $B$. In this situation the vertex functions
$\Gamma_\pm^{RR(RA)}(\br_1,\br_2)$ are, however, neither gauge- nor
translationally invariant. We define a gauge- and translationally
invariant vertex function by introducing a phase factor induced by the
vector potential ${\bf A}({\bf r})$ on a straight line connecting
$\br_1$ and $\br_2$,  
\be
\label{gauge}
\tilde{\Gamma}_\pm^{RR(RA)}(\br)= \exp\,[\,-i{e\over c}{\bf A}({\bf R}){\bf
  r}\,]\,\Gamma_\pm^{RR(RA)}(\br_1,\br_2)~, 
\ee 
where $\br =\br_1-\br_2$ and ${\bf R}=(\br_1+\br_2)/2$. After the
Fourier transformation, $\br\to\bp$ [note that $\bp$ has the meaning
of the kinematic rather than canonical momentum, in view of the
transformation (\ref{gauge})], we get then the following equation for
the dressed vertex:
\bea
\label{vertex}  
&& \tilde{\Gamma}_\pm^{RR(RA)}(\bp) = p_\pm/m +
4\sum_n (-1)^nG_{n\pm 1}^R(\ve+\omega) G_n^{R(A)}(\epsilon) \nonumber \\
&& \times \int{d^2p'\over(2\pi)^2} \tilde{W}(\bp-\bp')
e^{-l_B^2p'^2}L_n^1(2l_B^2p'^2)\tilde{\Gamma}_\pm^{RR(RA)}(\bp')~,
\eea
with $p_\pm=p_x\pm i p_y/ 2$. Using the asymptotic behavior of the Laguerre
polynomial $L_n^1(x)$ at $n,x\gg 1$, one can show that the following
replacement is justified \cite{vertex-note} in Eq.~(\ref{vertex}):
\be
(-1)^n e^{-l_B^2p'^2}L_n^1(2l_B^2p'^2) \longrightarrow
\delta(2l_B^2p'^2-4n)~.
\label{replace}
\ee
The sum over $n$ in Eq.~(\ref{vertex}) is then dominated by a narrow
band of width $\delta n/n\sim 1/\epsilon_F\tau_{s,0}$ around the Fermi
surface. Exploiting the SCBA condition $d/v_F\tau_{s,0}\ll 1$, we
finally reduce Eq.~(\ref{vertex}) to the form
\bea
\label{vertex-fs}
\tilde{\Gamma}_\pm^{RR(RA)}(\phi) &=& {v_F\over 2} e^{\pm i\phi} +
2\pi\nu_0 \Pi_\pm^{RR(RA)} 
\int{d\phi'\over 2\pi} \nonumber \\
&\times &
\tilde{W}(2k_F\sin{\phi-\phi'\over 2})
\tilde{\Gamma}_\pm^{RR(RA)}(\phi')~.
\eea
Therefore, the inclusion of the vertex correction results in a
replacement of $\Pi_\pm^{RR(RA)}$ in Eq.~(\ref{sigma2}) by
\be
\Pi_{\pm,{\rm tr}}^{RR(RA)}\equiv\left[\left(\Pi_\pm^{RR(RA)}\right)^{-1}
-(\tau_{s,0}^{-1}-\tau_{{\rm tr},0}^{-1})\right]^{-1}.
\label{PiTr}
\ee 
Evaluating Eq.~(\ref{sigma2}) with this substitution to first order in
$\delta$, we get the following result for the {\it ac} conductivity at
zero temperature, $T=0$:
\bea\label{corr1}
&&\sigma^{(1)}_\pm(\omega)=\sigma^{(D)}_\pm(\omega)\left\{
  1 - 2\delta\cos (2\pi\ve_F/\omega_c)  \right.\\
&&\times\left.\left[\frac{2\alpha_\pm^2}{\alpha_\pm^2+1} \frac{\sin(
      2\pi\omega/\omega_c)}{2\pi\omega/\omega_c}+
    \frac{3\alpha_\pm^2+1}{\alpha_\pm^2+1}
    \frac{\sin^2(\pi\omega/\omega_c)}{\alpha_\pm\pi\omega/
      \omega_c}\right]\right\}~,\nonumber 
\eea 
where $\alpha_\pm\equiv\tau_{{\rm tr},0}(\omega\pm\omega_c)$. Let us
stress that the single-particle time $\tau_{\rm s,0}$ enters
Eq.~(\ref{corr1}) only through the damping factor $\delta$; everywhere
else it has been replaced by the transport time $\tau_{\rm tr,0}$ due
to the vertex correction \cite{dc-note}.

Since the correction in Eq.~(\ref{corr1}) oscillates with $\ve_F$, it
becomes damped at finite $T$ by the factor $X/\sinh X$ with $X=2\pi^2
T/\omega_c$. If $T$ is higher than the Dingle temperature $T_D\equiv
1/2\pi\tau_{\rm s,0}$, the temperature smearing becomes the dominant
damping factor. In ultra-clean systems of the type used in the
experiments \cite{zudov,mani} the Dingle temperature is as low as
$T_D\sim 100\:{\rm mK}$, so that for characteristic measurement
temperatures $T\sim 1\:{\rm K}$ the first-order correction
(\ref{corr1}) will be completely suppressed. We will show, however,
that there exists a correction, oscillatory in $\omega/\omega_c$,
which is not affected by the temperature. To obtain it, we have to
extend our calculation to second order in $\delta$. Analyzing all
arising terms, we find that the required contribution is generated
only by the expansion of $\Pi_{\pm,{\rm tr}}^{RA}$, Eq.~(\ref{PiTr}),
with $\Pi_\pm^{RA}$ given by Eq.~(\ref{Pi}), to second order in
$\delta$, $(\Sigma^{(1)*}_\ve-\Sigma^{(1)}_{\ve+\omega})^2 \to
2\delta^2 \exp(-2\pi i\omega/\omega_c)$. Note that there is no need to
calculate $\Sigma_\ve$ to second order, neither to take $\Pi_\pm^{RR}$
into account, since the corresponding terms oscillate with $\ve$. We
thus get the following result for the leading quantum correction at
$T\gg T_D$:
\bea\label{corr2}
&& \sigma^{(2)}_\pm(\omega) = \sigma^{(D)}_\pm(\omega)\left\{1+2\delta^2
\right.\\
&&\times\left.  \left[\frac{\alpha_\pm^2(\alpha_\pm^2-3)}
    {(\alpha_\pm^2+1)^2}\cos\frac{2\pi\omega}{\omega_c} +
    \frac{\alpha_\pm(3\alpha_\pm^2-1)}{(\alpha_\pm^2+1)^2}
    \sin\frac{2\pi\omega}{\omega_c}\right]\right\}~.\nonumber 
\eea 

The regime which is most interesting theoretically and relevant
experimentally is that of long-range disorder, $\tau_{\rm
tr,0}/\tau_{\rm s,0}\gg 1$, and a classically strong magnetic field,
$\omega_c,\omega\gg \tau_{\rm tr,0}^{-1}$.  In this situation
$|\alpha_\pm| \gg 1$ and Eq.~(\ref{corr2}) acquires a remarkably
simple form (Fig.~2),

\be
\label{corr2-smooth}
\sigma^{(2)}_\pm(\omega) = \sigma^{(D)}_\pm(\omega)\, [\,1 + 2
e^{-2\pi/\omega_c\tau_{\rm s,0}}\cos(2\pi\omega/\omega_c)\,]~.
\ee

\begin{figure}
\narrowtext
\centerline{ {\epsfxsize=7.5cm{\epsfbox{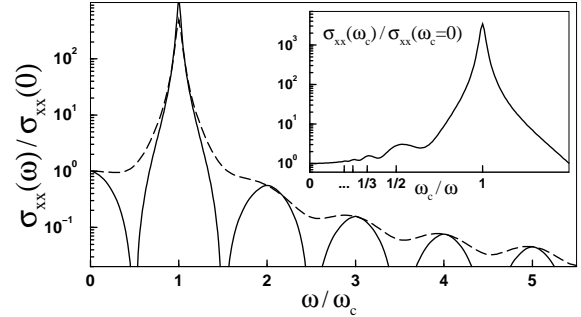}} }}
\caption{Magnetooscillations of the dynamical conductivity for a
system with   smooth disorder, $\tau_{\rm tr,0}/\tau_{\rm
s,0}=10$. Solid line: separated   LLs, $\omega_c\tau_{\rm
s,0}/\pi=3.25$; dashed line: overlapping LLs, $\omega_c\tau_{\rm
s,0}/\pi=1$. Inset: $\sigma_{xx}$ for fixed $\omega\tau_{\rm
s,0}/2\pi=1$ as a function of $\omega_c$.}
\label{fig2}
\end{figure}

We turn now to the opposite limit of fully separated LLs, when the LL
width $2\Gamma=2(2\omega_c/\pi\tau_{\rm s,0})^{1/2}$ \cite{raikh} is
small compared to $\omega_c$. This calculation can be done either by
using Eqs.~(\ref{sigma2}),(\ref{PiTr}) or directly in the LL
representation. We will only briefly present the main results
(illustrated in Fig.~2); details will be presented elsewhere
\cite{long}. The conductivity $\sigma_{xx}(\omega)$ is non-zero only
for $\omega$ in intervals $[M\omega_c-2\Gamma,\:M\omega_c+2\Gamma]$
with an integer $M$. At the center of the $M=1$ interval we find a CR
peak of height $\sigma_{xx}(\omega=\omega_c)= (e^2\nu_0
v_F^2/\pi\Gamma) \tau_{\rm tr,0}/\tau_{\rm s,0}$ and width $\sim
\Gamma \tau_{\rm s,0}/\tau_{\rm tr,0}$. All other peaks ($M\ne 1$) are
much smaller [in view of $\tau_{\rm tr,0}/\tau_{\rm s,0}\sim (k_F
d)^2\gg 1$]:
\be
\sigma_{xx}(\omega=M\omega_c)=\frac{4e^2 \nu_0 v_F^2 \Gamma }{3\pi\omega_c^2}
\frac{\tau_{\rm s,0}}{\tau_{\rm tr,0}}\frac{M^2+1}{(M^2-1)^2}~.  
\ee 
In fact, the whole dependence $\sigma_{xx}(\omega)$ can be described
as a Drude-type structure with a renormalized DOS,
$\nu(\ve)=\nu_0\tau_{\rm s,0}[\Gamma^2-(\ve-\ve_n)^2]^{1/2}$ for
$|\ve-\ve_n|\le \Gamma$, and the transport time $\tau_{\rm
tr}(\ve)=\tau_{\rm tr,0}\nu_0/\nu(\ve)$:
\bea \nonumber
\sigma_\pm(\omega)=\frac{e^2 v_F^2}{4\omega}\!\int\!
\frac{d\ve\,(f_\ve-f_{\ve+\omega})\,\nu(\ve)\,\tau_{\rm
    tr}^{-1}(\ve+\omega)}{[\tau^{-2}_{\rm tr}(\ve)+\tau^{-2}_{\rm
    tr}(\ve+\omega)]/2+(\omega\pm\omega_c)^2}~. 
\eea

Finally, we discuss the relation to the photoconductivity
oscillations. For this purpose, we present first a simple way to
understand the oscillatory structure in the {\it ac} conductivity
(\ref{corr2-smooth}). Since ${1\over 2}\sigma_{xx}(\omega)E_\omega^2$
is the power absorbed in a linearly polarized {\it ac} field
$E_\omega$, we have
\be
\label{sigma-qualit}
\sigma_{xx}(\omega) \sim \omega^2 \,V\,|M(\omega)|^2\, \langle
\nu(\ve)\nu(\ve+\omega)\rangle_\ve~,
\ee
where $M(\omega)$ is the dipole matrix element, and
$\langle\ldots\rangle_\ve$ denotes the energy averaging with the
corresponding Fermi factors [see Eq.~(\ref{sigma1})]. Using
Eq.~(\ref{dens}) for the DOS oscillations, we get
$\langle\nu(\ve)\nu(\ve+\omega)\rangle_\ve \simeq
\nu_0^2\,[\,1+2\delta^2\cos(2\pi\omega/\omega_c)\,]$ at $T\gg T_D$,
reproducing the oscillatory quantum correction in
Eq.~(\ref{corr2-smooth}). In a similar way, we now consider the
photoconductivity. The concentration of photo-excited electrons is
$n\sim [\,\sigma(\omega)/\omega\,]E_\omega^2\tau_{\rm in}$, where
$\tau_{\rm in}$ is the {\it inelastic} relaxation time, and we assumed
that the amplitude $E_\omega$ of the {\it ac} field is sufficiently
weak. A correction to the {\it dc} conductivity induced by these
electrons due to the energy dependence of the DOS can then be
estimated as
\bea
\label{photo}
\sigma_{\rm ph} &\sim&\omega\:(eE_{\omega}l_{\rm in})^2 \,V\,|M(\omega)|^2\,
\langle\nu(\ve)\nu'(\ve+\omega)\rangle_\ve
\nonumber \\
&\sim & -\frac{(eE_{\omega}l_{\rm in})^2}{\omega\omega_c} 
\,\sigma^{(D)}_{xx}(\omega)
\,e^{-2\pi/\omega_c\tau_{\rm s,0}}\sin{2\pi\omega\over\omega_c}~,
\eea
where $l_{\rm in}=v_F(\tau_{\rm tr, 0}\tau_{\rm
in})^{1/2}/\omega_c\tau_{\rm tr,0}$ is the inelastic length and we
used Eqs.~(\ref{corr2-smooth}), (\ref{sigma-qualit}) in the second
line. Note that the steady state to linear order in the radiation
power is reached only due to inelastic processes, leading to
$\sigma_{\rm ph}\propto\tau_{\rm in}$. We believe that
Eq.~(\ref{photo}) describes the leading contribution to $\sigma_{\rm
ph}$ induced by the LL quantization in the limit $\tau_{\rm in}\gg
\tau_{\rm s, 0}$. Let us stress that this contribution comes from an
oscillatory correction to the distribution function, as opposed to the
mechanism of Ref.~\cite{durst}, related to the effect of microwaves on
the collision integral.

Although Eq.~(\ref{photo}) agrees with the experiment as far as the
period \cite{zudov,mani} and the phase \cite{mani} of the oscillations
are concerned, there is a considerable disagreement in the damping of
oscillations at low $B$.  Specifically, our consideration predicts a
damping factor $\delta^2=e^{-2\pi/\omega_c\tau_{\rm s,0}}$ [same as in
the ac-conductivity, Eqs.~(\ref{corr2}),~(\ref{corr2-smooth})], so
that if the experimental data for the damping of the photoconductivity
oscillations are fitted to the form $e^{-\pi/\omega_c\tau_{\rm ph}}$,
one should find $\tau_{\rm ph}/\tau_{\rm s,0}={1\over 2}$. On the
other hand, the experiments yield much larger values, $\tau_{\rm  
ph}/\tau_{\rm s,0}\simeq 13\:{\rm ps}/2.5\:{\rm ps}=5.2$ \cite{zudov}
and $\tau_{\rm ph}/\tau_{\rm s,0}\simeq 18\:{\rm   ps}/11\:{\rm
ps}\simeq 1.6$ \cite{mani}. In other words, the photoconductivity
oscillations are observed at such low fields that the contribution due
to the above mechanism should be completely suppressed.

This suggests that, at least at lower fields, another mechanism, not
related directly to the LL quantization, might govern the observed
oscillatory photoconductivity.  A possible candidate is quasiclassical
memory effects. It has been shown that they may induce strong {\it dc}
magnetoresistance \cite{memory} and generate harmonics of the CR
\cite{ac-antidots} in models with smooth disorder and/or strong
scatterers.  It is thus natural that the memory effects induce also a
quasiclassical oscillatory contribution to the photoconductivity; work
in this direction is in progress \cite{long,aleiner}.

To summarize, we have studied the {\it ac} magnetoconductivity
$\sigma_{xx}(\omega)$ of a 2DEG with smooth disorder characteristic of
high-quality semiconductor structures. The interplay of Landau
quantization and disorder induces a contribution oscillating with
$\omega/\omega_c$ (the CR harmonics).  The effect is suppressed both
in the classical limit $\omega_c\tau_{\rm s,0}\to 0$ and in the clean
limit $\omega_c\tau_{\rm s,0}\to \infty$, and can be best observed in
the crossover range, $\omega_c\tau_{\rm s,0}\sim 1$. We have discussed
the relation to the recent experiments on photoconductivity of
ultra-clean samples \cite{zudov,mani}. A much stronger damping of our
result (\ref{corr2-smooth}) for weak $B$ suggests that another
(quasiclassical) mechanism may govern the observed oscillations
\cite{zudov,mani} in the low-field region.

We thank K.~von~Klitzing, R.~G.~Mani, J.~H.~Smet, and M.~A.~Zudov for
information about the experiments, and I.~V.~Gornyi and F.~von~Oppen
for stimulating discussions. This work was supported by the
Schwerpunktprogramm ``Quanten-Hall-Systeme'' and the SFB195 der
Deutschen Forschungsgemeinschaft, and by RFBR.

\end{multicols}

\vspace{-0.5cm}

\begin{references}

\vspace{-1.5cm}

\bibitem[*]{byline} Also at A.F.~Ioffe Physico-Technical
Institute, 194021 St.~Petersburg, Russia.

\bibitem[\dagger]{byline} Also at Petersburg Nuclear Physics
Institute, 188350 St.~Petersburg, Russia.

\bibitem{ando75} T.~Ando, J.~Phys. Soc. Japan {\bf 38}, 989 (1975).

\bibitem{afs} T.~Ando, A.B.~Fowler, and F.~Stern, Rev. Mod. Phys. {\bf
54}, 437 (1982).

\bibitem{abstreiter76} J.P.~Kotthaus, G.~Abstreiter, and J.F.~Koch,
Solid State Commun. {\bf 15}, 517 (1974); G.~Abstreiter,
J.P.~Kotthaus, J.F.~Koch, and G.~Dorda, Phys. Rev. B {\bf 14}, 2480
(1976).  

\bibitem{footAndo} The consideration of ultra-long-range disorder with
$d\gg l_B$ in \cite{ando75} is not justified since this condition
violates the applicability of SCBA.

\bibitem{zudov} M.A.~Zudov, R.R.~Du, J.A.~Simmons, and J.L.~Reno,
Phys. Rev. B {\bf 64}, 201311(R) (2001); M.A.~Zudov, R.R.~Du,
L.N.~Pfeiffer, and K.W.~West, Phys. Rev. Lett. {\bf 90}, 046807
(2003); C.L.~Yang {\it et al.}, cond-mat/0303472.

\bibitem{mani} R.G.~Mani, J.H.~Smet, K.~von~Klitzing,
V.~Narayanamurti, W.B.~Johnson, and V.~Umansky, Nature {\bf 420}, 646
(2002); cond-mat/0303034.

\bibitem{durst} A.C.~Durst, S.~Sachdev, N.~Read, and S.M.~Girvin,
cond-mat/0301569. 

\bibitem{anderson} P.W. Anderson and W.F. Brinkman, cond-mat/0302129.

\bibitem{shi} J.~Shi and X.C.~Xie, cond-mat/0302393.

\bibitem{ryzhii} V.I.~Ryzhii, Sov. Phys. Solid State {\bf 11}, 2078
(1970); V.I.~Ryzhii, R.A.Suris, and B.S.~Shchamkhalova,
Sov. Phys. Semicond. {\bf 20}, 1299 (1986). 

\bibitem{raikh} M.E.~Raikh and T.V.~Shahbazyan, Phys. Rev. B {\bf 47},
1522 (1993); B.~Laikhtman and E.L.~Altshuler, Ann. Phys. {\bf
232}, 332 (1994).

\bibitem{vertex-note} The function in Eq.~(\ref{replace}) has a width
$\delta p^\prime\sim n^{1/3}/l^2_B p^\prime\sim 1/n^{1/6}l_B$, which
is small compared to the width $\delta p^\prime\sim d^{-1}$ of the
rest of the integrand in Eq.~(\ref{replace}), in view of the SCBA
condition $d\ll l_B$.

\bibitem{dc-note} In the {\it dc} case, the results for $\sigma_{xx}$
and $\sigma_{xy}$ obtained in this way [A.D.~Mirlin, I.V.~Gornyi, and
F.~von~Oppen (unpublished)] confirm the form of Shubnikov-de Haas
oscillations in smooth disorder conjectured in P.T.~Coleridge,
R.~Stoner, and R.~Fletcher, Phys. Rev. B {\bf 39}, 1120 (1989).

\bibitem{long} I.A.~Dmitriev, A.D.~Mirlin, and D.G.~Polyakov, to be
published. 

\bibitem{memory} A.D.~Mirlin {\it et al}, Phys. Rev. Lett. {\bf 83},
2801 (1999); Phys. Rev. Lett. {\bf 87}, 126805 (2001); A.~Dmitriev,
M.~Dyakonov, and R.~Jullien, Phys. Rev. B {\bf 64}, 233321 (2001).

\bibitem{ac-antidots} D.G.~Polyakov, F.~Evers, and I.V.~Gornyi, 
Phys. Rev. B {\bf 65}, 125326 (2002). 

\bibitem{aleiner} I.L.~Aleiner, A.V.~Andreev, and B.L.~Altshuler,
private communication. 

\end{references}
\end{document}